# Nonlocal study of the near field radiative heat transfer between two n−doped semiconductors

F. Singer, Y. Ezzahri and K. Joulain*

Institut Pprime, Université de Poitiers-CNRS-ENSMA
2, Rue Pierre Brousse, Bâtiment B25, TSA 41105
86073 Poitiers Cedex 9, France.
* : karl.joulain@univ-poitiers.fr

## Abstract

We study in this work the near-field radiative heat transfer between two semi-infinite parallel planes of highly n-doped semiconductors. Using a nonlocal model of the dielectric permittivity, usually used for the case of metallic planes, we show that the radiative heat transfer coefficientsaturates as the separation distance is reduced for high doping concentration. These results replace the *1/d²* infinite divergence obtained in the local model case. Different features of the obtained results are shown to relate physically to the parameters of the materials, mainly the doping concentration and the plasmon frequency.

## I.   Introduction

Near field radiative heat transfer (NFRHT) had drawn a lot of attention in the past years due to the different physical phenomena attributed to it. In the near field, typically at distance shorter than Wien's thermal wavelength (10μm at temperature *T*=300K), physical phenomena such as the tunneling of the evanescent wavescontribute significantly to the transfer, changing by that the mechanisms of transfer with respect to the far-field [1−6]. The tunneling evanescent electromagnetic (EM) wavesdecay exponentially from the interfaces so that their contribution vanishes as long as the separation distance is larger than the wavelength. Their role in the NFRHT was first pointed out by Cravalho et al. [7] and Polder and Van Hove [8] in their pioneering work of studying the radiative heat transfer (RHT) in the near field. They have shown that when two bodies are approached, the radiative heat flux (RHF) between them increases significantly until reaching values of many orders higher than that between two blackbodies [4,6,9].This increase is a consequence of the new channels of transfer corresponding to modes of large wavevectors parallel to the surface by which heat transfer is enhanced, i.e. the tunneling evanescent EM waves. Following the work of Cravalho et al.[7] and Polder and Van Hove [8], many theoretical studies were carried out for systems consisting of two semi-infinite planeparallel solid surfaces. The studies aimed to calculate the NFRHT between the considered planes using a local dielectric permittivity function, as the optical response of the material was considered local, i.e. $\varepsilon = \varepsilon(\omega)$ where $\omega$ is the angular frequency of the EM wave [4,9,10−13]. The resultsshowed different behaviors according to the type of the considered material at ultra-short separation distances. For dielectrics, the NFRHT follows a $1/d^2$ law starting at distances as large as few hundreds of nm [4,6,8,12,13], where $d$ denotes the separation distance between the planes.For metals, the transfer seems to saturate at distances below the material skin depth and then diverges with a $1/d^2$ law at extremely small separation distance $d$ below 1 nm [5,12,14]. Following these theoretical predictions, some experimental studies were carried out to study the RHT between different bodies as the separation distance decreases [14−18]. The obtained results confirmed the enhancement of the RHF in the nanometer regime due to the tunneling of evanescent EM waves at the surfaces. Some of the experimental studies have roughly confirmed the $1/d^2$ law mostly at micrometric distances [16].

Physically, the $1/d^2$ diverging law as the separation distance $d$ is reduced, cannot be followed at extremely small distances as no heat transfer can become infinite; in addition to the fact thatat the atomic scale, the continuous behavior of matter does not exist andthe matter response necessarily changes for high spatial frequency. This leads to the need of a nonlocal description of the matter response as suggested by various authors [4,9,19], where the dielectric permittivity function will not be only frequency dependent but also wavevector dependent. For the case of dielectrics, Singer et al. [6] have recently shown that saturation of theradiative heat transfer coefficient (RHTC)between two semi-infinite parallel dielectric planes at short distances is achieved using a single-oscillator model in combination with the so-called hydrodynamic model of the dielectric permittivity [20,21], based on Halevi−Fuchs theory [22].For the case of two metallic planes, Chapuis et al. [5] studied few years ago the Lindhard−Mermin nonlocal dielectric permittivity function model and showed that the NFRHT saturated at distances of the order of the Thomas−Fermi length and also suppressed the $1/d^2$ divergence that occurred at extremely small distances.

In this article, we are interested in studying the case of highly doped semiconductors. To our knowledge, no nonlocal model of the dielectric permittivity for doped semiconductors has been studied and approved in the NFRHT study. We will carry our calculations by considering highly n-doped silicon (n−Si) planes. Doped Si is obtained when some of the Si atoms in the diamond crystal lattice are replaced by electron donors (As, P, Sb) or electron acceptors (B, Al, In, Ga), and it is thus called n-doped or p-doped Si, respectively. Due to these impurities, the doped silicon is capable of supporting surface plasmon polaritons in the near infrared range. The resonant angular frequency is thus found in the near infrared and its position depends on the nature of the doping (n or p) and the amount of dopants, or in other words, the doping concentration $N$. Doped Si is of great interest in different technology domains including the thermos-photovoltaic applications that are attracting a lot of attention and baring extensive research [23−38]; our near-field study of n-Si could be of great concern for the latter researches and could enhance their applications.

In the n-doped Si, the number of outer electrons acquired by the dopant exceeds the numbers of the silicon atoms. When the dopant concentration is so high that it is comparable to the effective density of states of Si, or the host semiconductor in general, the band of donor states overlap the bottom of the conduction band. It follows that when the concentration of electrons in the conduction band exceeds the effective density of states (in the conduction band), the Fermi energy lies within the conduction band and the highly n−doped Si is called a degenerate semiconductor. This results in changing the properties of Si (or the host semiconductor in general) to resemble those of metals. These facts drive us to represent the optical properties of this material using the dielectric permittivity models used for metals and eventually applying them in the study of the RHTC between two highly n−doped Si planes.

Our system consists of two semi-infinite parallel solid planes of temperatures $T_1$ and $T_2$, respectively. They are considered to be separated by a vacuum gap of width $d$ (Fig.1). The RHF is obtained by considering the fluctuational−dissipation theorem that describes the EM field in the near field using Maxwell's equations and was first suggested by Rytov [12,39]. The detailed explanation of this theorem and the detailed derivation of the RHF are not given here as they are well presented in many references [1,2,4,8,40−42]. Therefore, we will only recall the final expressions of the RHF $\phi(T,d)$ and the RHTC $h_{rad}(T,d)$ [4,6,42].

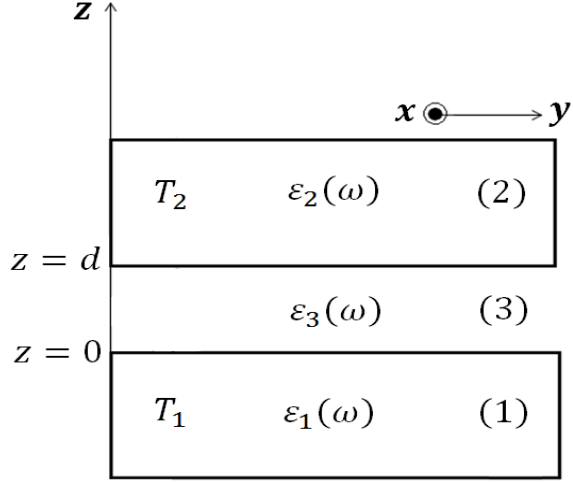

**Figure 1:** Two parallel semi-infinite material planes separated by a vacuum gap of width $d$.

$$\begin{cases} \phi(T,d) = h_{rad}(T,d)\delta T \\ h_{rad}(T,d) = \sum_{\alpha=S,P} \int_0^{+\infty} d\omega [h_{prop}^\alpha(T,d,\omega) + h_{evan}^\alpha(T,d,\omega)] \\ h_{prop}(T,d,\omega) = h^0(T,\omega) \times \int_0^{k_0} \frac{KdK}{k_0^2} \frac{(1-|r_{31}^\alpha|^2)(1-|r_{32}^\alpha|^2)}{|1-r_{31}^\alpha r_{32}^\alpha e^{2i\gamma_3 d}|^2} \\ h_{evan}(T,d,\omega) = h^0(T,\omega) \times \int_{k_0}^{+\infty} \frac{KdK}{k_0^2} \frac{4Im(r_{31}^\alpha)Im(r_{32}^\alpha)e^{2i\gamma_3 d}}{|1-r_{31}^\alpha r_{32}^\alpha e^{2i\gamma_3 d}|^2} \end{cases} \quad (1)$$

$\delta T = T_1 - T_2$ such that $\delta T/T_1 \ll 1$. $k_0 = \omega/c$, $\omega$ is the wave angular frequency and $c$ is the speed of light in vacuum, $K$ and $\gamma_3 = \sqrt{k_0^2 - K^2}$ are the wavevector components parallel and normal to the surface in vacuum, respectively. $r_{31}^\alpha$ and $r_{32}^\alpha$ represent the reflection factors for the EM waves of polarization $\alpha = s, p$ incident from medium 3 and reflected on medium 1 and 2, respectively. $h^0(T,\omega)$ is the derivative of the blackbody specific intensity of radiation with respect to temperature (Planck's law), see Eq. (2).

$$h^0(T,\omega) = \frac{\hbar\omega^3}{4\pi^2 c^2} \frac{\hbar\omega}{k_B T^2} \left[2\sinh\left(\frac{\hbar\omega}{2k_B T}\right)\right]^{-2} \quad (2)$$

where $\hbar = 1.054571 \times 10^{-34} J.s$ is the reduced Planck constant and $k_B = 1.380648 \times 10^{-23} J.K^{-1}$ is Boltzmann constant. It is clear in Eqs. (1) that the RHTC is the sum of the contributions of propagative $(K < k_0)$ and evanescent $(K > k_0)$ waves of $s$ and $p$ polarizations. The general expressions of the reflection factors (Eqs. 3) [4,6,42] depend on the surface impedances $Z_m^\alpha$ between media 3 and $m$ defined as the ratio of the parallel component of the electric field on the parallel component of the magnetic field. Their general expressions are given in Eqs. 4 [5,6,42,43].

$$\begin{cases} r_{3m}^p = \dfrac{\gamma_3 - \varepsilon_3\,\omega\, Z_m^p}{\gamma_3 + \varepsilon_3\,\omega\, Z_m^p} \\ r_{3m}^S = \dfrac{c^2\gamma_3\, Z_m^S - \omega}{c^2\gamma_3\, Z_m^S + \omega} \end{cases} \quad (3)$$

$$\begin{cases} Z_m^p = \dfrac{2i}{\pi\omega}\displaystyle\int_0^{+\infty}\dfrac{dq}{k^2}\left[\dfrac{q^2}{\varepsilon_t(k,\omega) - (ck/\omega)^2} + \dfrac{K^2}{\varepsilon_l(k,\omega)}\right] \\ Z_m^S = \dfrac{2i}{\pi\omega}\displaystyle\int_0^{+\infty}\dfrac{dq}{\varepsilon_t(k,\omega) - (ck/\omega)^2} \end{cases} \quad (4)$$

$k^2 = q^2 + K^2$ where $k$ is the total wavevector, $q$ and $K$ are the perpendicular and the parallel wavevector components, respectively. $\varepsilon_t(k,\omega)$ and $\varepsilon_l(k,\omega)$ denote the transverse and the longitudinal components of the dielectric permittivity. Eqs. (4) show that in the general case, the dielectric permittivity function is dependent on $\omega$ and $k$. In the local case, by assuming the medium to be isotropic, $\varepsilon_t(k,\omega) = \varepsilon_l(k,\omega) = \varepsilon(\omega)$. Eqs.(4) are thus simplified and by substituting them in the expressions of the reflection factors (Eq. (3)) we end up with the very known Fresnel reflection factors Eqs. (5).

$$\begin{cases} r_{3m}^p = \dfrac{\varepsilon_m\gamma_3 - \varepsilon_3\gamma_m}{\varepsilon_m\gamma_3 + \varepsilon_3\gamma_m} \\ r_{3m}^S = \dfrac{\gamma_3 - \gamma_m}{\gamma_3 + \gamma_m} \end{cases} \quad (5)$$

where $\varepsilon_m = \varepsilon(\omega)$ is the dielectric permittivity of the medium m and $\gamma_m = \sqrt{\varepsilon_m k_0^2 - K^2}$ is the normal wavevector component in the medium $m$. To calculate the RHTC between two semi-infinite parallel planes of n−Si, we use the Drude local model of the dielectric permittivity, usually used for metallic planes[4,5,31,42,44]. We considered the average temperature of the n−Si system to be $T$=300K ($T_1 = 300.5K$ and $T_1 = 299.5K$. The results obtained show that for any considered doping concentration, the contributions of the propagative EM waves of *s* and *p* polarizations and the contribution of the evanescent EM waves of *s*-polarization saturated. The contribution of the evanescent EM waves of *p*-polarization diverged non-physically at extremely small distances as *1/d²*; these results are similar to those obtained in previous work and recalled in the introduction [5,12,14].

In the following section we will repeat the RHTC study using Lindhard−Mermin nonlocal model of the dielectric permittivity. In section three we will present the results and their interpretations andthe last section will be devoted to our conclusions.

## II. Theory of the nonlocal model of the dielectric permittivity: Lindhard−Mermin nonlocal model

In this section, we repeat the calculation of the RHTC between two n−doped Si semi-infinite parallel planes of average temperature $T$=300K by considering a nonlocal model of the dielectric permittivity. This model differs from the local one by accounting for spatial dispersion in the medium so that the dielectric permittivity function is of frequency and wavevector dependence, i.e.$\varepsilon = \varepsilon(\omega, k)$. As we mentioned in the introduction, due the metallic-like properties of the highly n−doped Si, we consider in our calculations the Lindhard−Mermin nonlocal model of the dielectric permittivity usually applied for metals [5,40,45−49]. This model describes the optical properties of a semi-infinite free electron gas. It was extended by Mermin [47] from the longitudinal dielectric constant derived by Lindhard [50] to obtain finite-electron-lifetime generalized equations. Mermin used the relaxation−time approximation to account for the collisions in the electron gas; the longitudinal term expression obtained is combined with the transverse term derived by Ford and Weber [48,49] to form the nonlocal dielectric model used in our study.

This model is the classical nonlocal model of the dielectric permittivity for metalsand it was shown by Chapuis et al. [5]that using it in the study of the RHTC between two semi-infinite parallel planes of Al lead to saturation along with different features that were well explained, analytically and physically[5,42]. It follows that the validity of this nonlocal model in the case of highly n-doped Si system is supported not only by the physical characteristics of the material that lead to using this model in the first place, but bythe physical significance of the upcoming results as well.

Lindhard−Mermin nonlocal model of thedielectricpermittivity is composed of longitudinal and transversal termsgiven by Eqs. (6) [5,42]:

$$\begin{cases} \varepsilon_{LM}^l(\omega, z) = \varepsilon_b + \dfrac{3\,\varepsilon_b \omega_p^2}{\omega + i\nu} \dfrac{u^2}{\{\omega + i\nu[f_l(z,u)/f_l(z,0)]\}} \\ \varepsilon_{LM}^t(\omega, z) = \varepsilon_b - \dfrac{\varepsilon_b \omega_p^2}{\omega^2(\omega + i\nu)} \{\omega[f_t(z,u) - 3z^2 f_l(z,u)] + i\nu[f_t(z,0) - 3z^2 f_l(z,0)]\} \end{cases} \quad (6)$$

where the detailed expressions of $f_l(z,u), f_t(z,u), f_l(z,0)$ and $f_t(z,0)$ are given in the references [5,42]. $z = k/2k_F$ and $u = (\omega + i\nu)/kv_F$, $k_F = m^*v_F/\hbar$ is the Fermi wavevector, $v_F = \left(\frac{2k_BT}{m^*}\log[N/N_c]\right)^{1/2}$ is the Fermi velocity and $\nu = Ne^2\rho/m^*$ is the losses factor. $N$ denotes the doping concentration, $N_c$ is the density of carriers at the bottom of the conduction band (for $T$=300K, $N_c \approx 3.5 \times 10^{18} cm^{-3}$). $m^* = 0.27 \times m_0$ is the electron effective mass, $m_0 = 9.109 \times 10^{-31} Kg$ is the free electron mass, and $e = 1.602 \times 10^{-19} C$ is the electron charge. $\varepsilon_b = 11.7$ is the static permittivity. The expression of $v_F$ is deduced from the relation $E_F = \frac{1}{2}m^*v_F^2$, taking into account that $E_F \approx k_BT\log[N/N_c]$ for the n−doped semiconductors. The validity of the model implies that $v_F > 0$ which is equivalent to the imperative condition $N > N_c$ that should be always satisfied. $\rho$ is the electric resistivity of the doped Si[23,51] and $\omega_p$ is the plasma frequency given by $\omega_p^2 = Ne^2/m^*\varepsilon_b\varepsilon_0$, where $\varepsilon_0 = 8.854 \times 10^{-12} F.m^{-1}$ is the vacuum permittivity.

It is important to highlight the fact that the main difference between applying the Lindhard−Mermin model for metals and applying it for n−doped Si is the dependence of the Fermi velocity $v_F$ on the doping concentration $N$ in the case of Si, where for metals it is a constant [5,42]. The dependence of the dielectric permittivity and the different parameters of the material on the doping concentration $N$ allows us to predict a direct dependence of the properties of the obtained results on $N$.

The reflection factors and the surface impedances needed to calculate the RHTC are given by the general Eqs. (3) and Eqs. (4), respectively. By substituting these equations in the RHTC equations (Eqs. (1)), we are able to calculate the different contributions to the RHTC between two semi-infinite parallel n−Si planes of average temperature $T$=300K as the separation distance $d$ decreases, for the cases where $N = 10^{19} cm^{-3}, 10^{20} cm^{-3}$ and $10^{21} cm^{-3}$.

### III. Results and discussions

In Fig. 2 we present the plots of the different contributions to the RTHC for $N = 10^{20} cm^{-3}$ using Lindhard−Mermin nonlocal model of the dielectric permittivity, and the total contributions to the RHTC using the same model for three different values of the doping concentration.

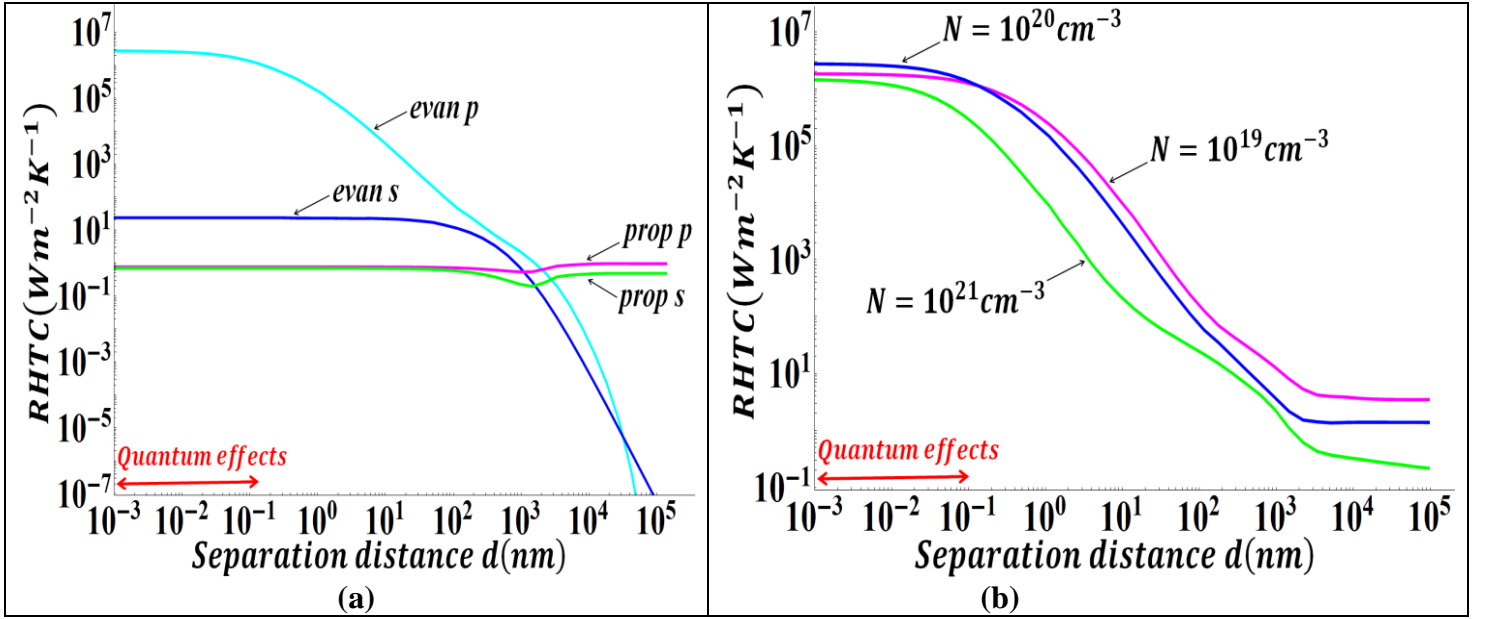

**Figure 2: (a)** Variation of the contributions of the evanescent and the propagative EM waves of *s* and *p* polarizations to the RHTC as function of the distance between two semi-infinite n−doped Si parallel planes of doping concentration $N = 10^{20}$ and average temperature *T*=300K, for the nonlocal model case. **(b)** Variation of the total radiative heat transfer coefficient (summation of the contributions of the evanescent and propagative EM waves of *s* and *p* polarizations) as function of the distance between two semi-infinite n−doped Si parallel planes of doping concentration $N = 10^{19}\,\text{cm}^{-3}, 10^{20}\,\text{cm}^{-3}$ and $10^{21}\,\text{cm}^{-3}$, and average temperature *T*=300K for the nonlocal model case. The two-headed arrow highlights the distance range in which quantum effects come into play.

In Fig. 2 (a) we observe that all the contributions to the RHTC saturated at short distances for $N = 10^{20}$. Similar plots are obtained for the cases where $N = 10^{19}\,\text{cm}^{-3}$ and $N = 10^{21}\,\text{cm}^{-3}$ except that the saturation values are different. In Fig.2 (b) we observe that the RHTC saturated at short distances for the three cases of the different doping concentrations. To analyze the different contributions to the RHTC and their effects, we will start with those of the propagative EM waves of *s* and *p* polarizations. Thecontributions of these waves almost do not depend on the separation distance *d*(in contrary to the contributions of the evanescent EM waves of *s* and *p* polarizations) and they attain relatively small values at short distances, rendering them with negligible effects to the RHTC at these distances scales. The saturation of the contribution of the evanescent EM waves of *s*-polarization which is well interpreted in the local case [5,42] is completely similar to the nonlocal case;this contribution is not affected by the nonlocality. The nonlocality affected only the contribution of the evanescent waves of *p*-

polarization;itsinfinite divergence is replaced by a physical saturation of finite value at short distances.By presenting the plots of the RHTC for distances as short as $10^{-12} m$, we show that saturation is mathematically obtained at extremely short distances where the modeling presented here more likely ceases to be valid. Indeed, below 5 Å quantum effects dominate; i.e. when the separation distances are around the typical atomic distances, quantum effects appear especially for metals (and eventually highly doped semiconductors) where electrons are the dominant heat carriers [52−54].It follows thatthese quantum effects should be accounted for starting from distances of the order of the lattice constant ($a \approx 5.43$Å for Si)as to obtain a physically correct and complete study of the RHT [55]. To analyze the saturation results obtained, we plot in Fig. 3 the contributions of the evanescent EM waves of *p*-polarization in the local and the nonlocal cases within the "physical" distance range of our study (minimum distance $\approx 1$Å). At distances of the order of $10 nm$ the curves of the nonlocal model corresponding to the cases where $N = 10^{19} \text{cm}^{-3}$ and $10^{20} \text{cm}^{-3}$ deviate from the diverging curves of the local model. The deviation of these curves is followed by aslowincreasing rate of their values. The curves thensaturateat values two orders of magnitude smaller than those of the diverging curves of the local model. The deviation of the curve of the nonlocal model of $N = 10^{21} \text{cm}^{-3}$ from that of the diverging curve of the local model takes place at $d \approx 1 nm$. The curve did not reach the saturation value at $d = 1$Å, but from Fig. 2b we observe that saturation takes place at shorter distances. We notice that the curves of the nonlocal model tend to saturate starting from distances of the order of the Thomas−Fermi length $l_{TF} = \vartheta_F/\omega_p$ [5,41]; $l_{TF}(N = 10^{19} \text{ cm}^{-3}) \approx 1.86 \text{ nm}$, $l_{TF}(N = 10^{20} \text{ cm}^{-3}) \approx 1.05 \text{ nm}$, and $l_{TF}(N = 10^{21} \text{ cm}^{-3}) \approx 0.43 \text{ nm}$. Since for metals, and eventually for highly doped materials that exhibit metallic-like characteristics, nonlocal effects appear at distances smaller than the Thomas−Fermi length [56], it is expected that the saturation starts to take place at distances of the order of $l_{TF}$; a result attained before for the metallic systems [5,42].In Fig. 3 we also notice that the values (almost equal to the saturation values) reached by the curves of the cases $N = 10^{19} cm^{-1}$ and $N = 10^{20} cm^{-1}$ at extremely short distances, are equal and larger than that of the curve of the case $N = 10^{21} cm^{-1}$. For all distances larger than few angstroms, the value of the RHTCfor the nonlocal model decrease as *N* increases; a trend shown by the results of the local model for all distances. To explain this result, we refer to the interpretations given in references [42,44].It was shown that for the case of two n-doped Si planes, the RHT is enhanced significantly due to the excitation of surface plasmon-polaritons, as in the case of metals. The excitation frequency of these surface waves was found to be $\omega_{plasmon} \approx \omega_p$. Due to the dependence of the plasma frequency, and eventually the plasmon frequency on the doping level *N*, the value of $\omega_p$ will change as *N* changes. This implies that as the value of $\omega_p$ vary within the Planck range, the corresponding spectrum shows a similar behavior as Planck spectrum; i.e. the radiative spectrum will increase in values as $\omega_p$ increases, until reaching a maximum, after which it decreases in magnitude. This explains the decrease in the values of the RHTC as *N* decreases, in the local model and the nonlocal model cases.

The above results show that taking into account the nonlocal effects in the dielectric permittivity equations allows representing correctly the optical response at short distances and lead finally to the physical saturation of the RHTC between the highly n−doped Si planes.

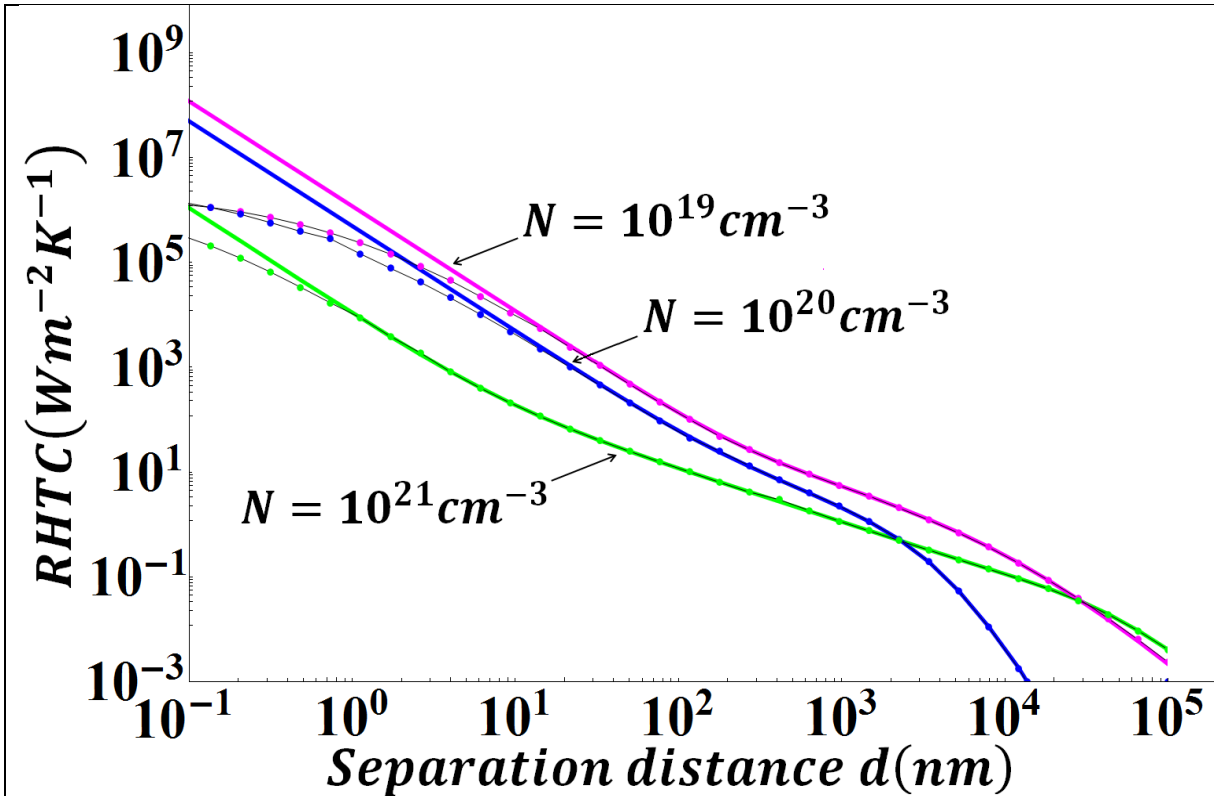

**Figure 3:** Variation of the contribution of the evanescent EM waves of *p*-polarization to the RHTC as function of the distance between two semi-infinite n-doped Si parallel planes of doping concentration $N = 10^{19} cm^{-3}, 10^{20} cm^{-3}$ and $10^{21} cm^{-3}$ and average temperature *T*=300K, for the local and the nonlocal model cases. For any *N*, the graph of the nonlocal model tend to saturate starting from distances of the order of the Thomas−Fermi length; $l_{TF}(N = 10^{19} cm^{-3}) \approx 1.86\ nm$, $l_{TF}(N = 10^{20} cm^{-3}) \approx 1.05\ nm$, and $l_{TF}(N = 10^{21} cm^{-3}) \approx 0.43\ nm$.

An important factor to study is the transmission coefficient of the *p*-polarized evanescent EM waves $4(Im(r_{31}^P))^2 e^{2i\gamma_3 d}/|1-(r_{31}^P)^2 e^{2i\gamma_3 d}|^2$. In Fig.4 we plot its variation in the ($\omega$,*K*) plane at different distances for the local and the nonlocal models, for the doping concentration $N = 10^{20} cm^{-3}$ and average temperature *T* =300K.

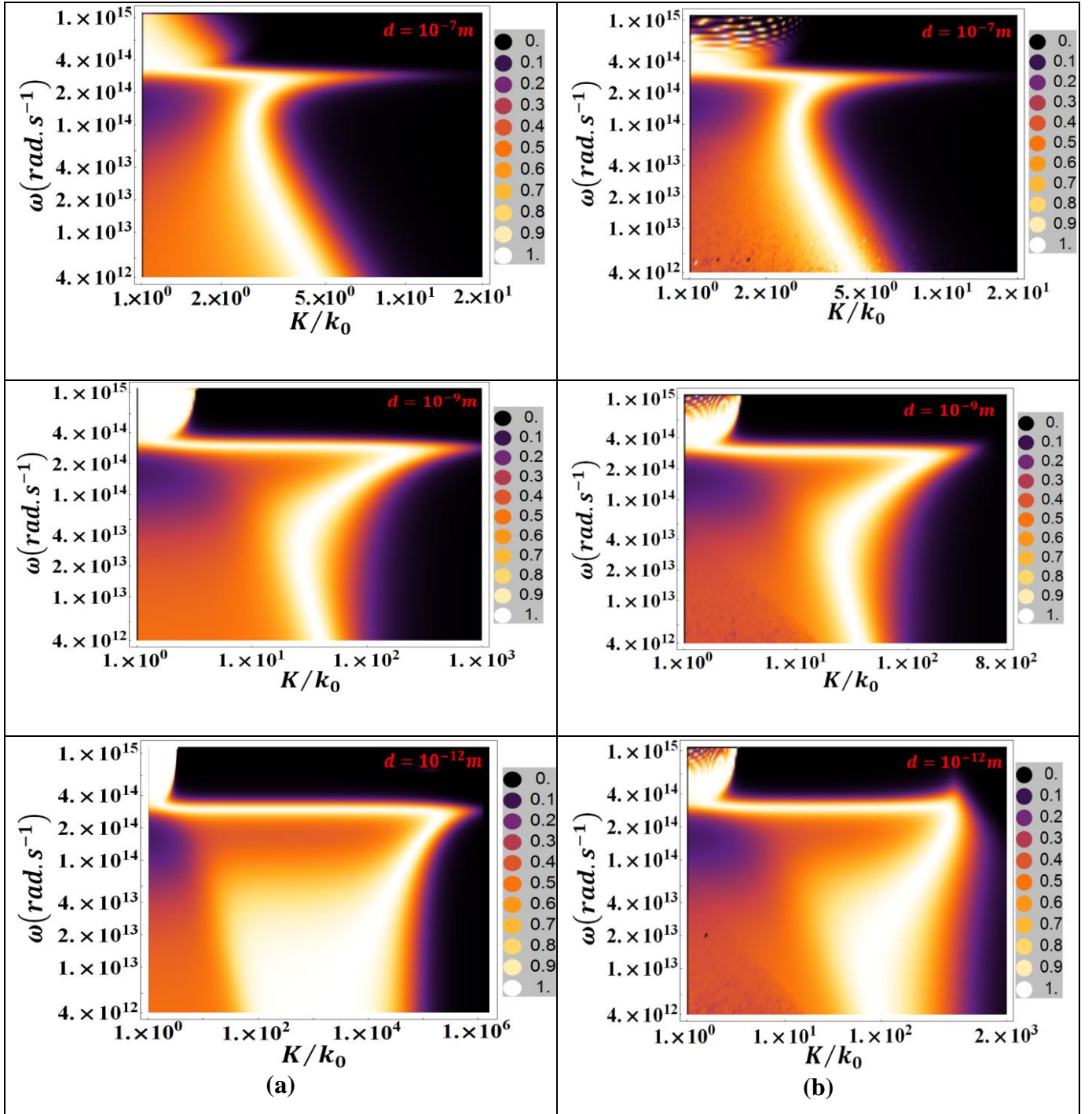

**Figure 4:** Plot of the transmission coefficient $4(Im(r_{31}^P))^2 e^{2i\gamma_3 d}/|1-(r_{31}^P)^2 e^{2i\gamma_3 d}|^2$ of the *p*-polarized evanescent EM waves in the plane *(ω,K)* for the local model case **(a)** and the nonlocal model case **(b)** for the doping level $N = 10^{20} cm^{-3}$ and average temperature *T*=300K and at different separation distances *d*.

We notice from the curves of the local model in Fig. 4(a) that at any *d*, the transmission factor increases in magnitude as *K* increases. As the distance decreases the maximum is reached at larger *K* values. This implies that more modes are able to couple well to the transmission as *K* increases and *d* decreases leading to the increases of the RHT.This is thus consistent with the results shown in Fig. 3 for the infinite increase of the RHTC as the separation distance decreases. From the curves of the nonlocal model in Fig. 4 (b), we notice that the plot of the transmission coefficient at $d = 10^{-7} m$ is quite similar to that of the local model case; at large distances, i.e. in the far-field regime, the nonlocal effects are negligible and the results obtained in the local and the nonlocal casesare the same. This is consistent with the results shown in Fig. 3 where we observe that the curves of the local and the nonlocal models overlap for all distances larger than few nanometers.For shorter separation distances,the transmission factor of the nonlocal model case increases as *K* increases; in all cases the range of *K* values covered by the spectrum is shorter than that covered by the spectrum of the local model case. This is consistent with the result shown in Fig. 3 where at distance of order of $1 nm$ the curves of the nonlocal model increase in a rate which is much slower than those of the local model curves and they acquire smaller values than those of the local model curves. At extremely short distance, the transmission coefficient of the nonlocal model acquires a cutoff after which the values decrease gradually and tend to zero where no modes are able to be transmitted. This cutoff takes place at *K*of the order of the inverse of the Thomas−Fermi length $K_{cuttoff} = 1/l_{TF} \approx 9.4 \times 10^8 m^{-1}$. This is consistent with the results obtained for the RHTC plots where theirtendency to saturate starts to take place at distances of the order of the Thomas−Fermi length. Therefore, the results obtained for the RHTC and the supporting results of the transmission coefficients spectrum emphasize the fact that the values of wavevectors larger than $K_F$ are not screened by the electron plasma. This leads to the decrease in the value of the imaginary part of the reflection factor and eventually to the decay of the transmission coefficient; the divergence of the RHT is thus limited. For the other doping concentrations, similar spectra of thetransmission coefficient are obtained, leading eventually tosimilar observations and arguments. In other words, as the value of the doping concentration decreases, the *K* value at which the cutoff takes place decreases, and in all cases it is of the order of the inverse of the Thomas−Fermi length.

## IV. Conclusions

In this work we studied the RHTC between two n−doped Si semi-infinite parallel planes as the distance between them decreases. The key-idea in this study is using the dielectric permittivity models of metals due to the fact that the highly doped Si acquires metallic-like properties. Since the usage of a local model of the dielectric permittivity leads to the infinite nonphysical *1/d²* divergence of the RHTC at extremely small distances, we suggested using a nonlocal model of the dielectric permittivity. In the second part of the paper we calculated the RHTC between the n−doped Si planes using the Lindhard−Mermin nonlocal dielectric model of the dielectric permittivity. We showed that for all *N*, the contribution of the evanescent *p*-polarized EM waves saturates as the distance decreases, in contrary to the divergence obtained in the local model case. The saturation in each case starts to take place at distances of the order of the Thomas−Fermi screening length $l_{TF}$, which is consistent with the fact that for metals and materials with metallic-like properties, the nonlocal effects appear at distances smaller than $l_{TF}$. We also studied the plot of the transmission factor in the plane (*ω,K*) for the local and the nonlocal cases at distances $d = 10^{-7}m, 10^{-9}m$ and $10^{-12}m$ for the case where $N = 10^{20} \text{cm}^{-3}$. In the local case the values of the transmission factor increase as *K* increases while the curves of the nonlocal case showed saturation at *K* values of the order of the inverse of the Thomas-Fermi length for short distances, supporting by this the related obtained results of the RHTC

**References**

1. J. J. Loomis and H. J. Maris, *Phys. Rev. B***50**, 18517- 18 524 (1994).
2. J.-P. Mulet, K. Joulain, R. Carminati, and J.-J. Greffet, *Microscale Therm. Eng.***6**, 209–222 (2002).
3. A. I. Volokitin and B. N. J. Persson; *Phys. Rev. B***69**, 045417 (2004).
4. K. Joulain, J.-P. Mulet, F. Marquier, R. Carminati, and J.-J. Greffet, *Surf. Sci. Rep.***57**, 59-112 (2005).
5. P. O. Chapuis, S. Volz, C. Henkel, K. Joulain and J.-J. Greffet; *Phys. Rev.***B** 77, 035431 (2008).
6. F. Singer, Y. Ezzahri and K. Joulain, J. Quantitative Spectroscopy and Radiative Transfer **154**, 55–62 (2015).
7. E.G. Carvalho, C.L.Tien, and R.P. Caren, *J. Heat Transfer***89**, 351-357 (1967).
8. D. Polder and M. Van Hove; *Phys. Rev. B***4**, 3303-3314 (1971).
9. A. Volokitin and B. Persson, *Rev. Mod. Phys.***79**, 1291-1329 (2007).
10. S. M. Rytov et al; *Principles of Statistical Radiophysics "Springer, New York"*, Vol. 3 (1989).
11. C. J. Fu and Z. M. Zhang; *Int. J. Heat Mass Transfer***49**, 1703-1718 (2006).
12. Z. M. Zhang; *Nano/Microscale Heat Transfer "McGraw-Hill, New York"* (2007).
13. *S. Basu,* Z. M. Zhang, and C. J. Fu; *Int. J. Energy Res.***33**, 1203-1232 (2009).
14. C. M. Hargreaves, *Phys. Lett.***30A**, 491 (1969).
15. G. A. Domoto, R. F. Boehm, and C. L. Tien, *J. Heat Transfer***92**, 412-416 (1970).
16. A. Kittel, W. Müller-Hirsch, J. Parisi, S.A. Biehs, D. Reddig, and M. Holthaus, *Phys. Rev. Lett.***95**, 224301 (2005).
17. L. Hu, A. Narayanaswamy, X. Chen and G. Chen, *Appl. Phys. Lett.***92**, 133106 (2008).
18. E. Rousseau et al, *Nat. Photonics***3**, 514-517 (2009).
19. C. Henkel and K. Joulain, *Appl. Phys. B***84**, 61-68 (2006).

20. N. W. Ashcroft and N. D. Mermin, *"Solid State Physics"*, Holt, Rinehart and Winston, New York, (1976).
21. J. J. Hopfield and D. G. Thomas, *Phys. Rev.* **132**, 563 (1963).
22. P. Halevi and R. Fuchs, *J. Phys. C: Solid State Phys.* **17**, 3869 (1984).
23. P.J. Hesketh, J.N. Zemel, B. Gebhart, *Nature (London)* **324,** 549 (1986).
24. A. Treatise, Semiconductors and Semimetals ed. by *R. K. WILLARDSON* and *A. BEER,* Vol. **24** (1987).
25. P.J. Hesketh, J.N. Zemel, B. Gebhart, *Phys. Rev. B* **37**, 10795 (1988).
26. P.J. Hesketh, J.N. Zemel, B. Gebhart, *Phys. Rev. B* **37**, 10803 (1988).
27. G.Bomchil, A. Halimaoui and **R.** Herino, *Microelectronic Eng.* **8**, 293 (1988).
28. H. Ehsani et al.,*"characteristics of degenerately doped silicon for spectral control in thermophotovoltaic systems".* Conference on thermophotovoltaic generation of electricity, Colorado Springs, CO (United States), Jul 1995.
29. S. Hava, J. Ivri, and M. Auslender, *J. Appl. Phys.* **85**, 7893 (1999).
30. M.U. Pralle et al.,*Appl. Phys. Lett.* **81,** 4685 (2002).
31. F. Marquier , K. Joulain, J.P. Mulet, R. Carminati, J.J. Greffet , *Opt. Commun.* 237, 379 (2004).
32. R. Soref, *IEEE J. Sel. Top. inQuan. Electr.* **12**, 1678 (2006).
33. M. Ahmadi, H. Houg Lau, R. Ismail, V. Arora,*Microelectronics J.* **40**, 547 (2009).
34. J. Cruz-Campa et al.,*Sol. Energ. Mat. Sol. C.* **95**, 551–558 (2011).
35. L. M. Fraas and L.D. Partain, *"Solar Cells and their Applications",*John& Wiley Sons, second edition (2010).
36. Y. Schiele, S. Fahr, S. Joos, G. Hahn and B. Terheiden, **"***Study on boron emitter formation by BBR₃ diffusion for n-type Si solar cell applications",* Preprint to the 28th EU-PVSEC, Paris 2013.
37. S. Boriskina, H. Ghasemi and G. Chen, *Mater. Today* **16**, 375-386 (2013).
38. E. Nefzauoi, PhD thesis report *"Conception et optimisation de sources thermiques cohérentes pour applications thermo-photovoltaïques"*,(2013).
39. *S. Rytov, Sov. Phys. JETP 6, 130-140 (1958).*
40. J. Pendry „*J. Phys. Condens. Matter* **11**, 6621-6633 (1999).
41. A. Volokitin and B. Persson, *Phys. Rev. B* **63**, 205404 (2001).
42. F. Singer, PhD thesis report *"Influence of the nonlocal effects on the near-field radiative heat transfer",* (2014).

43. K .L . Kliewer and R . Fuchs, Theory of dynamical properties of dielectric surfaces, Advance in Chemical Physics, Vol. XXVII, ed. I. Prigogine, Stuart A. Rice by John Wiley& Sons. (1974).
44. E. Rousseau, M. Laroche, and J.-J. Greffet, *Appl. Phys. Lett.* **95**, 231913 (2009).
45. K .L . Kliewer and R . Fuchs, *Phys. Rev.* **185**, 905-913 (1969).
46. K .L . Kliewer and R . Fuchs, *Phys. Rev.B* **3**, 2270-2278 (1970).
47. N. D. Mermin, *Phys. Rev. B* **1**, 2362 (1970).
48. G. W. Ford and W. H. Weber, *Phys. Rep.* **113**, 195 (1984). We note a mistake in formula 2.29 corrected, for example, in Ref. 45.
49. R. Esquivel and V. B. Svetovoy, *Phys. Rev. A* **69**, 062102 (2004).
50. J. Lindhard, Kgl. Danske Videnskab. Selskab, *Mat.-Fys. Medd.* **28**, 8 (1954).
51. J. Irvin, *Bell Sys.t Tech. J.* **XLI**, 387 (1961).
52. H. Duan , A. I. Fernández-Domínguez, M. Bosman, S. A. Maier and J. K. W. Yang, *Nano Lett.*, **12**, 1683 (2012).
53. J.A. Scholl, A.I. Koh and J.A. Dionne, *Nature,* **483**, 421 (2012).
54. F. Javier Garcia de Abajo, *Nature,* **483**, 417 (2012).
55. Y. Ezzahri and K. Joulain,*Phys. Rev. B* **90**, 115433 (2014).
56. P. Ben-Abdallah and K. Joulain, *Phys. Rev. B* **82**, 121419, (2010).